\documentclass[twocolumn,showpacs,preprintnumbers,amsmath,amssymb]{revtex4}
\usepackage{pifont}
\usepackage{tipa}
\usepackage{graphicx}
\usepackage{dcolumn}
\usepackage{slashbox}
\usepackage{bm}
\usepackage{multirow}

\begin{document}
\title{$D^\star_{s1}(2700)^\pm$ and $D^\star_{sJ}(2860)^\pm$ revisited within the $^3P_0$ model}

\author{Ling Yuan\footnote{leyou@shu.edu.cn}, Bing Chen
and Ailin Zhang\footnote{corresponding author:
zhangal@staff.shu.edu.cn}} \affiliation{Department of Physics,
Shanghai University, Shanghai 200444, China}
\begin{abstract}
The strong decays of $D^\star_{s1}(2700)^\pm$ and $D^\star_{sJ}(2860)^\pm$ are investigated within the $^3P_0$ model. It is found that the interpretation of these two states depends on the mixing schemes and the ways of choices of the harmonic oscillator parameter $\beta$. If $D^\star_{s1}(2700)^\pm$ and $D^\star_{sJ}(2860)^\pm$ are two pure states, $D^\star_{s1}(2700)^\pm$ seems impossibly the $2^3S_1$ $D_s$, but may be the $1^3D_1$ $D_s$. $D^\star_{sJ}(2860)^\pm$ may be the $1^3D_3$. If there is mixing between the $2^3S_1$ and $1^3D_1$, $D^\star_{s1}(2700)^\pm$ may be the mixed $1^-$ state with a small mixing angle in the case of a special $\beta$ for each meson, and $D^\star_{sJ}(2860)^\pm$ is the orthogonal partner of $D^\star_{s1}(2700)^\pm$; $D^\star_{s1}(2700)^\pm$ may also be the mixed $1^-$ state with a large mixing angle based on a universal $\beta$ for all mesons, and $D^\star_{sJ}(2860)^\pm$ seems impossibly the orthogonal partner of $D^\star_{s1}(2700)^\pm$. Other uncertainties related to the choices of constituent quark masses and phase spaces are also explored.
\end{abstract}

\maketitle

\section{INTRODUCTION} \label{sec1}
In the zoo of the heavy-light $D_s$ mesons, the properties of radially excited $2S$ and orbitally excited $1D$ states have been
explored for a long time. However, no such higher excited $D_s$ state has been definitely established. The observation of $D^\star_{s1}(2700)^\pm$ and $D^\star_{sJ}(2860)^\pm$ has stimulated much more interest in these higher excited states.

$D^\star_{s1}(2700)^\pm$ was first observed by Belle~\cite{belle1} in
$$B^+\to \bar D^0D_{s1}\to\bar D^0D^0K^+$$
with mass $\emph{M}=2715\pm 11^{+11}_{-14}$ MeV and width $\Gamma=115\pm 20^{+36}_{-32}$ MeV. $D^\star_{s1}(2700)^\pm$ was also observed by Babar in both $DK$ and $D^*K$ channels~\cite{babar1}. This state is included in PDG10~\cite{pdg1} with mass $\emph{M}=2709^{+9}_{-6}$ MeV, $J^P=1^-$ and width $\Gamma=125\pm30$ MeV. The branching ratio $\frac{{\cal B}(D_{s1}(2700)^+ \to D^*K)}{{\cal B}(D_{s1}(2700)^+ \to D K)} =0.91 \pm  0.13_{stat} \pm  0.12_{syst}$ has also been measured.

$D^\star_{sJ}(2860)^\pm$ was first reported by BaBar~\cite{babar2} in
$$D_{sJ}(2860)^+\to D^0K^+,~D^+K^0_s$$
with mass $M=2856.6\pm 1.5(stat)\pm 5.0(syst)$ and width $\Gamma=48\pm 7(stat)\pm10(syst)$ MeV. It was observed again in both $DK$ and $D^*K$ channels~\cite{babar1}. This state is included in PDG10~\cite{pdg1} with mass $M=2862\pm2^+5_-2$ MeV, width $\Gamma=48\pm 3\pm 6$ MeV and branching ratio $\frac{{\cal B}(D_{sJ}(2860)^+ \to D^*K)}{{\cal B}(D_{sJ}(2860)^+ \to D K)} =1.10 \pm 0.15_{stat} \pm 0.19_{syst}$. Its $J^P$ has not been measured or assigned. The observation of $D_{sJ}(2860)\to D^*K$ rules out the possibility as a $0^+$ state since a $1^3P_0$ $D_s$ is forbidden to decay into $D^*K$, and this state is supposed to have spin-parity: $1^-, 2^+, 3^-,\cdots$.

$D^\star_{s1}(2700)^\pm$ and $D^\star_{sJ}(2860)^\pm$ have been explained within some models. $D^\star_{s1}(2700)^\pm$ was identified
with the first radial excitation of $D_s^\ast(2112)^\pm$~\cite{bozhang2007,colangelo2008,bingchen2009,zhangal,zhangal1}, or the
orbitally excited D-wave $1^3D_1$ $D_s$~\cite{bozhang2007,zhangal1}, or the mixture of them~\cite{close2007,d.m.li,zhangal}. The interpretation of $D_{sJ}^\ast(2860)^\pm$ as the $2^3P_0$ $D_s$~\cite{close2007, beveren2006} was ruled out, and $D_{sJ}^\ast(2860)^\pm$ is interpreted as the $1^3D_3$ $D_s$~\cite{colangelo2006,bozhang2007,zhao,bingchen2009,zhangal,zhangal1}. However, theoretical
interpretations of these states in different assignments are not completely consistent with experiments either on their spectrum or on their decay properties. Where to place these two excited states? Obviously, the observations of $D^\star_{s1}(2700)^\pm$, $D^\star_{sJ}(2860)^\pm$ and some other resonances~\cite{babar3} bring us some puzzles. The study of the properties of strong decays is believed a good way to identify new observed states.

Among models for strong decays, the $^3P_0$ quark-pair-creation model has been employed successfully to evaluate the OZI-allowed strong decays of both mesons and baryons~\cite{micu1969,yaouanc1,yaouanc2,kogerler,stancu,barnes1,barnes2}. Thorough understanding of the success of the $^3P_0$ model has also been investigated~\cite{isgur,kokoski,stancu1,capstick,capstick1,geiger,ackleh}. Within the $^3P_0$ model, $D^\star_{s1}(2700)^\pm$ and $D^\star_{sJ}(2860)^\pm$ have been analyzed~\cite{bozhang2007,close2007,d.m.li}. However, there are different assignments and conclusions to these two states in these references. Especially, some theoretical predictions of the branching ratios $\frac{{\cal B}(D_{s1}(2700)^+ \to D^*K)}{{\cal B}(D_{s1}(2700)^+ \to D K)}$ and $\frac{{\cal B}(D_{sJ}(2860)^+ \to D^*K)}{{\cal B}(D_{sJ}(2860)^+ \to D K)}$ are not consistent with experiments, and the predicted mixing angle $\theta$ is different from each other. Therefore, it would be interesting to study the uncertainties relevant to the interpretations of $D^\star_{s1}(2700)^\pm$ and $D^\star_{sJ}(2860)^\pm$ within the $^3P_0$ model.

It is believed that the P-wave $D$ and $D_s$ mesons have been established~\cite{pdg1}. In the P-wave multiplets of heavy-light mesons, the $^3P_1$ and $^1P_1$ states may mix with each other, so the observed $J^P=1^+$ states should be the mixed ones. The mixing scheme has been explored and the mixing angle has been determined in Refs.~\cite{godfrey,close2005,zhao}. Similarly, the radially excited $2^3S_1$ and the orbitally excited $1^3D_1$ may mix with each other through some mechanism, which is known as the ``excited-vector-meson puzzle"~\cite{kokoski}. This mixing will complex our understanding of the higher excited $J^P=1^-$ states. Therefore, it will be useful to find out how large the mixing effect on the explanation of the excited states is. For example, to see whether the theoretical interpretation of the $J^P=1^-$ states is mixing scheme dependent or not, or to determine how large the mixing angle is.

In this paper, $D^\star_{s1}(2700)^\pm$ and $D^\star_{sJ}(2860)^\pm$ are re-studied within the $^3P_0$ model. The paper is organized as follows. In Sec.II, the strong decay widths and branching ratios $\Gamma(D^{\star}K)/\Gamma(DK)$ of $D^\star_{s1}(2700)^\pm$ and $D^\star_{sJ}(2860)^\pm$ are evaluated in different cases. Finally, we present our conclusions and discussions in Sec.III.

\section{Strong decays of $D^\star_{s1}(2700)^\pm$ and $D^\star_{sJ}(2860)^\pm$}

Since different assignments to $D^\star_{s1}(2700)^\pm$ and $D^\star_{sJ}(2860)^\pm$ have been suggested in literature, evaluations of their two-body open-flavor strong decays are made according to these assignments. In the conventional quark model, $D^\star_{s1}(2700)^\pm$ and $D^\star_{sJ}(2860)^\pm$ may be pure $n^{2S+1}L_J$ states, or the mixed states of $n^{2S+1}L_J$ states. If the $2^3S_1$ and $1^3D_1$ $D_s$ (same $J^P$ and similar masses) mix with each other, the physically observed states~\cite{close2007,d.m.li,zhangal} should be the mixed $1^-$ ones. In the mixed case, the two orthogonal partners are denoted as~\cite{zhangal}
\begin{eqnarray}\label{mixing}
|(SD)_1\rangle_L=cos\theta|2^3S_1 \rangle-sin\theta|1^3D_1\rangle\\\nonumber
|(SD)_1\rangle_R=sin\theta|2^3S_1 \rangle+cos\theta|1^3D_1\rangle,\\\nonumber
\end{eqnarray}
where $\theta$ is the mixing angle. In our evaluation, $D^\star_{s1}(2700)^\pm$ and $D^\star_{sJ}(2860)^\pm$ are first assumed as pure states, and subsequently assumed mixed states.

Within the $^3P_0$ model, the evaluation of the strong decays widths (formula in Ref.~\cite{bozhang2007} is adopted) involves some parameters: the strength of quark pair creation from the vacuum $\gamma$, the $\beta$ value in the simple harmonic oscillator (SHO) wave functions and the constituent quark masses. $\gamma$ represents the probability that a quark-antiquark pair is created from the vacuum, and it is supposed as a universal parameter. All the partial widths are proportional to $\gamma^2$. $\beta$ in the SHO wave function is the harmonic oscillator strength parameter, which can be fixed to reproduce the realistic root mean square radius of the SHO wave function. There are often two ways for choices of $\beta$. One way is to determine $\beta$ individually for each meson~\cite{close2005,d.m.li}, and the other way is to choose $\beta$ universal for all mesons~\cite{geiger,barnes3,zhangal}. Our investigation indicates that the ways of choices of $\beta$ play an important role in the evaluation of the strong decays. Therefore, our study of the decay widths is presented in the next two subsections in these two different ways, respectively. To extract the flavor coefficient, the ideal meson wave functions $\eta=\frac{1}{\sqrt{6}}(u\bar{u}+d\bar{d}-2s\bar{s})$ and $\omega=\frac{1}{\sqrt{3}}(u\bar{u}+d\bar{d}+s\bar{s})$ are used.

Other parameters are given as follows. $\gamma=6.25$~\cite{d.m.li} is employed, which leads also to the right total decay width of $D^\star_{s2}(2573)$. The masses of relevant pure mesons are taken from the 2010 Particle Data Group (PDG)~\cite{pdg1} or Ref.~\cite{isgur}. The constituent quarks masses are taken to be $m_u=m_d=0.22$ GeV, $m_s=0.42$ GeV and $m_c=1.628$ GeV~\cite{godfrey}.

\subsection{Strong decays with special $\beta$ for each meson}

In this subsection, $\beta$ is chosen as special value for each meson as that in Ref.~\cite{close2005}. $\beta$ values and relevant meson masses used for our evaluation of $D^\star_{s1}(2700)^\pm$ and $D_{sJ}(2860)\pm$ are listed in Table I.

\begin{table*}
\begin{tabular}{cccccccccccc}
\hline
\hline
States& $D_{s1}[2^3S_1]$ & $D_{s1}[1^3D_1]$ & $D_{sJ}[1^3D_3]$   &D  &K  &$D^\star$ &$D_s$&$D_s^\star$ &$\eta$ &$K^\star$&$\omega$\\
Mass(MeV) & 2709    & 2709    & 2862    &1865    &494     &2007    &1968    &2112    &547     &892     &783     \\
$\beta$(GeV)[21]& 0.35& 0.34& 0.34& 0.43& 0.46& 0.37& 0.52& 0.45& 0.48& 0.32& 0.36\\
\hline\hline
\end{tabular}
\caption{Meson masses and $\beta$ values used for the calculation of $D^\star_{s1}(2700)^\pm$ and $D^\star_{sJ}(2860)^\pm$} \label{table-2}
\end{table*}

In the case without mixing, we present the widths for all possible decay channels and the branching fraction ratio $\Gamma(D^{\star}K)/\Gamma(DK)$ in Table II, where the relativistic phase space is employed. As a pure $2^3S_1$ state, the predicted total decay width of $D^\star_{s1}(2700)^\pm$ is about half of the experimental data, but the predicted branching fraction ratio $\Gamma(D^{\star}K)/\Gamma(DK)$ is almost five times of the observed one. Since many systematic uncertainties in the branching fraction ratio are canceled out, the information extracted from the branching fraction ratio is often more reasonable. However, when the uncertainties explored in subsection C are taken into account, the branching fraction ratio may also has a large uncertainty. From Table. II, $D^\star_{s1}(2700)^\pm$ seems impossibly the pure $2^3S_1$. If $D^\star_{s1}(2700)^\pm$ is identified with a pure $1^3D_1$ state, its total decay width is approximately the same as experiment data but its branching fraction ratio $\Gamma(D^{\star}K)/\Gamma(DK)$ is half of the experiment. That is to say, $D_{s1}(2700)$ may be identified with the $1^3D_1$ with a large uncertainty.

\begin{table*}
\begin{tabular}{ccccccccc}
\hline\hline
Mode & $D^{\star}K$ & $DK$ & $D^\star_s\eta$ & $D_s\eta$ &$DK^{\star}$ & $D_s\omega$ &$\Gamma_{totall}$ & $\Gamma(D^{\star}K)/\Gamma(DK)$\\
\hline
$D_{s1}(2700)[2^3S_1]$    & 41.4    & 9.4    & 2.0    & 2.0    & -    & -    & 54.8    & 4.4\\
$D_{s1}(2700)[1^3D_1]$    & 39.1    & 93.8   & 2.0    & 16.7   & -    & -    & 151.6   & 0.42\\
$D_{sJ}(2860)[1^3D_3]$    & 22.7    & 32.8   & 0.7    & 1.9    & 2.1  &0.1   & 60.3    & 0.69\\
\hline\hline
\end{tabular}
\caption{Decay widths (MeV) and branching fraction ratio for $D^\star_{s1}(2700)^\pm$ and $D^\star_{sJ}(2860)^\pm$} \label{table-2}
\end{table*}

If $D^\star_{sJ}(2860)^\pm$ is the $1^3D_3$ $D_s$, their possible decay widths and branching fraction ratio $\Gamma(D^{\star}K)/\Gamma(DK)$ are also presented in Table II. Its total decay width is approximately consistent with experiment ($\Gamma=48\pm7$ MeV), and its predicted branching ratio $\Gamma(D^{\star}K)/\Gamma(DK)$ is about half of the observed result. Similarly, $D^\star_{sJ}(2860)^\pm$ may be a $1^3D_3$ $D_s$ with a large uncertainty.

In the mixing scheme indicated by Eq.~(\ref{mixing}), both partial widths and branching ratio $\Gamma(D^{\star}K)/\Gamma(DK)$ depend on the mixing angle $\theta$. Therefore, the mixing angle $\theta$ can be determined through comparing of theoretical results with experiments. The dependence of the branching ratio on the mixing angle $\theta$ is shown in Fig. 1(a), and the dependence of the decay widths on the mixing angle $\theta$ is shown in Fig. 1(b). If $D^\star_{s1}(2700)^\pm$ is identified with the state $|(SD)_1\rangle_L$, to obtain comparable results with experiments in both decay widths and branching ratio, the mixing angle $\theta$ is fixed at $-88^\circ\leq\theta\leq -76^\circ$. At this mixing angle, the total width of $D^\star_{s1}(2700)^\pm$ is found around: $\Gamma\simeq(111\pm 1)$ MeV. The fixed mixing angle $\theta$ is different from those determined in Ref.~\cite{d.m.li} ($1.12 \leq\theta\leq 1.38$) and Ref.~\cite{close2007} ($\theta\approx -0.5$). Obviously, the suggestion that $D^\star_{s1}(2700)^\pm$ is the mixture of the $2^3S_1$ and $1^3D_1$ $D_s$ is mostly favored.

\begin{figure}
\begin{center}
\includegraphics[height=4.3cm]{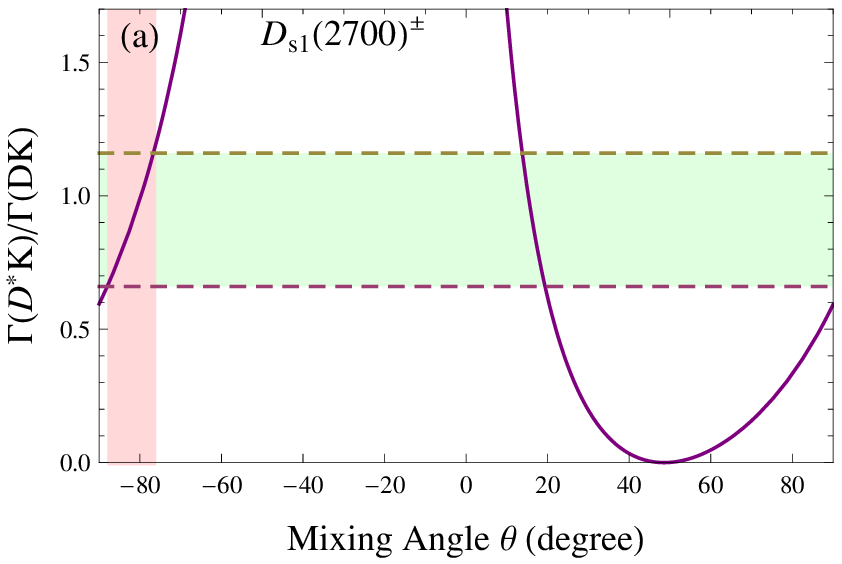}
\includegraphics[height=4.3cm]{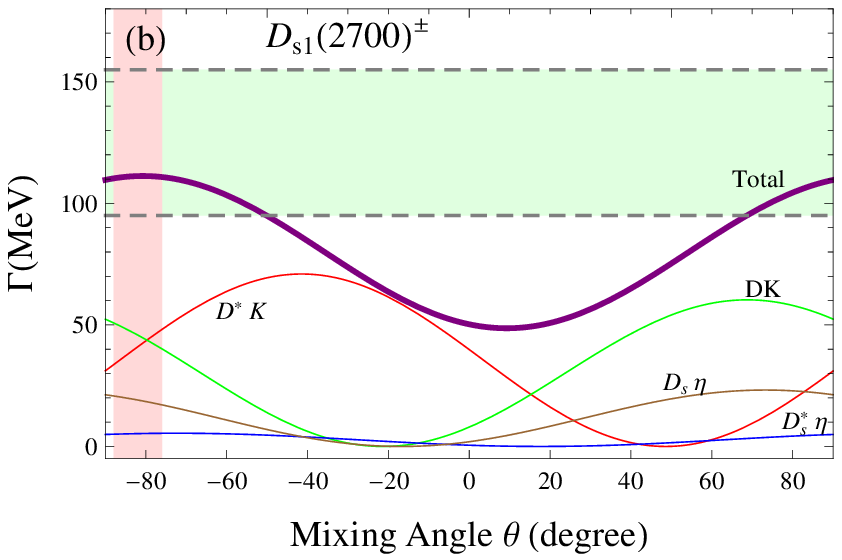}
\includegraphics[height=4.3cm]{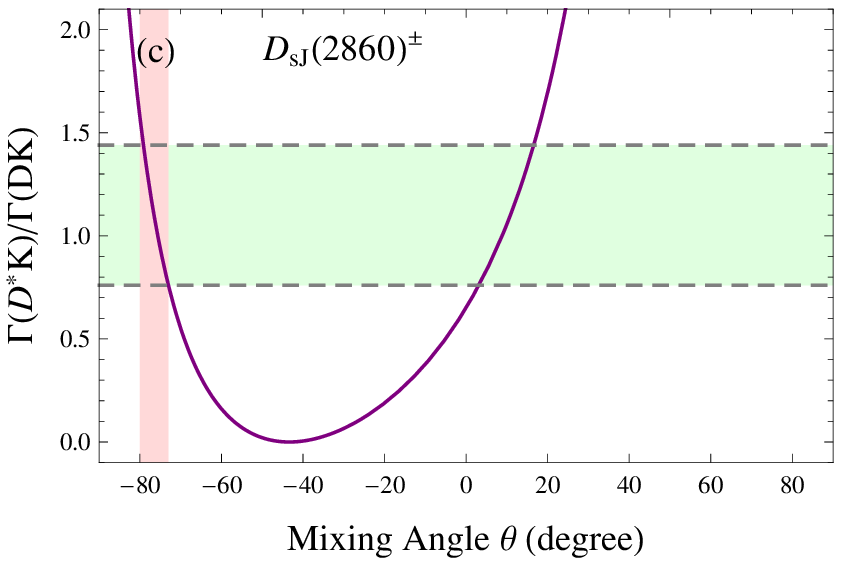}
\includegraphics[height=4.3cm]{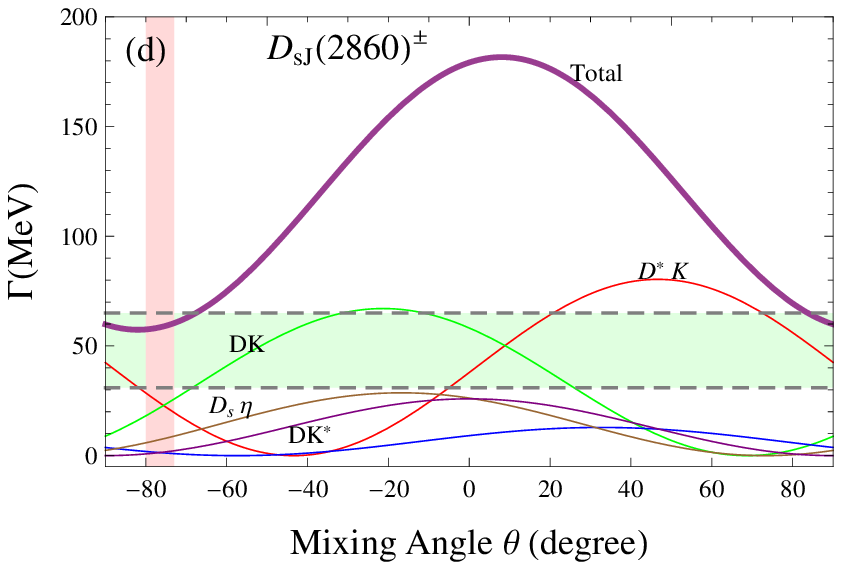}
\caption{Decay widths and $\Gamma(D^{\star}K)/\Gamma(DK)$ of $D^\star_{s1}(2700)^\pm$ and $D^\star_{sJ}(2860)^\pm$ versus $\theta$. The horizontal dashed lines indicate the upper and lower limits of the PDG data.}
\end{center}
\end{figure}

Following the same process, we evaluate the total decay width, partial decay widths and branching ratio $\Gamma(D^{\star}K)/\Gamma(DK)$ of $|(SD)_1\rangle_R$. The branching ratio and decay widths dependence on the mixing angle $\theta$ are shown in Fig. 1(c) and Fig. 1(d), respectively. From these two figures, it is found that the decay widths and branching ratio $\Gamma(D^{\star}K)/\Gamma(DK)$ of $D^\star_{sJ}(2860)^\pm$ can be properly reproduced by the $|(SD)_1\rangle_R$ with $-80^\circ\leq \theta \leq -73^\circ$, which is almost the same one determined by $D^\star_{s1}(2700)^\pm$. In other words, experimental data supports strongly the assignment of $D^\star_{sJ}(2860)^\pm$ as the $|(SD)_1\rangle_R$.

In summary, $D^\star_{s1}(2700)^\pm$ seems impossible the pure $2^3S_1$ $D_s$, but may be interpreted as the $1^3D_1$. $D^\star_{sJ}(2860)^\pm$ may be interpreted as the pure $1^3D_3$. In the mixing scheme, $D^\star_{s1}(2700)^\pm$ is very possibly the mixed $D_s$ of $2^3S_1$ and $1^3D_1$, and $D^\star_{sJ}(2860)^\pm$ is very possibly the orthogonal partner of $D^\star_{s1}(2700)^\pm$. Experiments are interpreted quite well with a large mixing angle $\theta\approx -80^\circ$.

\subsection{Strong decays with universal $\beta$ for all mesons}

Once $\beta$ is chosen as a universal parameter for all mesons~\cite{geiger,barnes3,zhangal}, $\beta=0.38$ GeV is fixed in our evaluation ($\beta$ is usually preferred at $\beta=(0.35-0.42)$ GeV).

To proceed the analysis, $D^\star_{s1}(2700)^\pm$ and $D^\star_{sJ}(2860)^\pm$ are studied first without taking into account the mixing effect. If $D^\star_{s1}(2700)^\pm$ is the pure $2^3S_1$ or $1^3D_1$ $D_s$, and $D^\star_{sJ}(2860)^\pm$ is the pure $1^3D_3$ $D_s$, their decay widths are presented in Table. III.
\begin{table*}
\begin{center}
\begin{tabular}{ccccccccc}
\hline\hline
\backslashbox{Assignment}{Mode}
& $D^{\star}K$ & $DK$ & $D^\star_s\eta$ & $D_s\eta$ &$DK^{\star}$ & $D_s\omega$ &$\Gamma_{totall}$ & $\Gamma(D^{\star}K)/\Gamma(DK)$\\
\hline
$D_{s1}(2700)[2^3S_1]$   & 84.8   & 31.8   & 10.5   & 20.4   & -    & -    & 147.5   & 2.67\\
$D_{s1}(2700)[1^3D_1]$   & 43.7   & 89.9   &3.3    & 30.0   & -    & -     & 166.8   & 0.48\\
$D_{sJ}(2860)[1^3D_3]$   & 23.6   & 42.2   & 1.5   & 6.2    & 1.4  &0.2    & 75.1    & 0.56\\
\hline\hline
\end{tabular}
\caption{Decay widths (MeV) and branching fraction ratios for $D^\star_{s1}(2700)^\pm$ and $D_{sJ}(2860)$} \label{table-2}
\end{center}
\end{table*}

The numerical results without mixing are different from those based on individual $\beta$ for each meson, but conclusions to $D^\star_{s1}(2700)^\pm$ and $D^\star_{sJ}(2860)^\pm$ are the same. If $D^\star_{s1}(2700)^\pm$ is the pure $2^3S_1$, its total decay width is a little larger than the experiment and the predicted branching ratio is about three times of the experiment. $D^\star_{s1}(2700)^\pm$ seems impossible the pure $2^3S_1$ $D_s$. If $D^\star_{s1}(2700)^\pm$ is a pure $1^3D_1$ state, its total decay width is a little larger than experiment data but its branching fraction ratio $\Gamma(D^{\star}K)/\Gamma(DK)$ is half of the experiment. $D_{s1}(2700)$ may be identified with the $1^3D_1$ with a large uncertainty.
If $D^\star_{sJ}(2860)^\pm$ is identified with the $1^3D_3$, its total decay width is a little larger than the experiment and the predicted branching ratio $\Gamma(D^{\star}K)/\Gamma(DK)$ is about half of the observed one. $D^\star_{sJ}(2860)^\pm$ may be the $1^3D_3$ $D_s$ with a large uncertainty.

Under the mixing scheme Eq. (\ref{mixing}), the decay widths and branching ratios $\Gamma(D^{\star}K)/\Gamma(DK)$ of $|(SD)_1\rangle_L$ and $|(SD)_1\rangle_R$ are calculated. To determine the mixing angle $\theta$, the decay widths and branching ratios $\Gamma(D^{\star}K)/\Gamma(DK)$ on the mixing angle $\theta$ is shown in Fig. 2. Once $D^\star_{s1}(2700)^\pm$ is regarded as the $|(SD)_1\rangle_L$, the mixing angle is fixed at two different places: {\bf $-90^\circ\leq\theta\leq -85^\circ$} and $12^\circ\leq\theta\leq 21^\circ$ by comparing the decay widths and branching ratio $\Gamma(D^{\star}K)/\Gamma(DK)$ with experiments. The $-90^\circ\leq\theta\leq -85^\circ$ may be excluded by the study of the spectrum of these two states~\cite{zhangal}, and the $12^\circ\leq\theta\leq 21^\circ$ is similar to that in Ref.~\cite{zhangal}.

The fixed mixing angle $\theta$ is different from that based on individual $\beta$ for each meson. In addition, $D^\star_{sJ}(2860)^\pm$ is difficult to be identified with the $|(SD)_1\rangle_R$, the orthogonal partner of $D^\star_{s1}(2700)^\pm$. At the mixing angles $\theta$, the branching ratio $\Gamma(D^{\star}K)/\Gamma(DK)$ of $D^\star_{sJ}(2860)^\pm$ is larger than the observed value. Furthermore, the total decay width ($>150$ MeV) of $D^\star_{sJ}(2860)^\pm$ is much broader than the observed $\Gamma=48\pm 3\pm 6$ MeV. In Ref.~\cite{zhangal}, the inconsistence of the theoretical evaluation with experimental data was explained in another way by the introduction of a new $D_{sJ}(2850)^\pm$. In short, within the mixing scheme, $D^\star_{s1}(2700)^\pm$ can be identified with the $|(SD)_1\rangle_L$ with a small mixing angle $12^\circ\leq\theta\leq 21^\circ$, but $D^\star_{sJ}(2860)^\pm$ is hard to be interpreted as the orthogonal partner $|(SD)_1\rangle_R$.
\begin{figure}\label{pic4}
\begin{center}
\includegraphics[height=4.3cm]{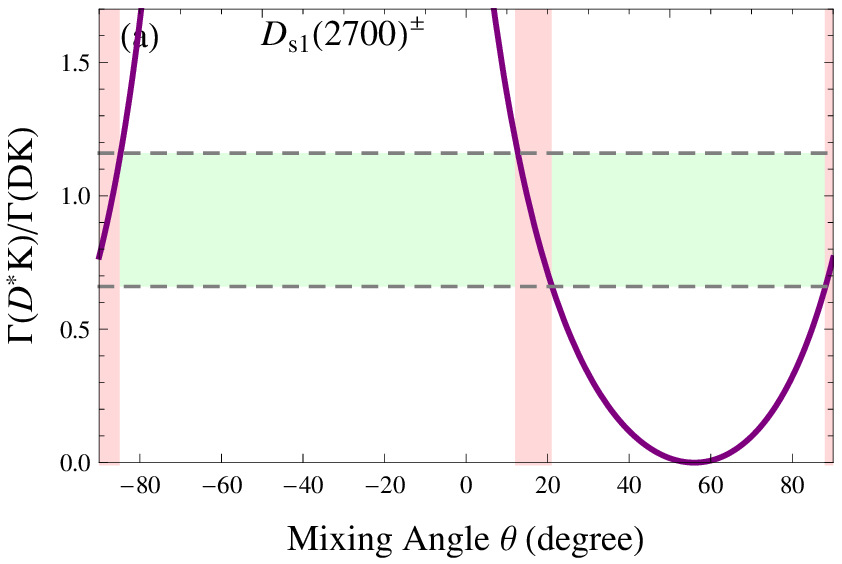}
\includegraphics[height=4.3cm]{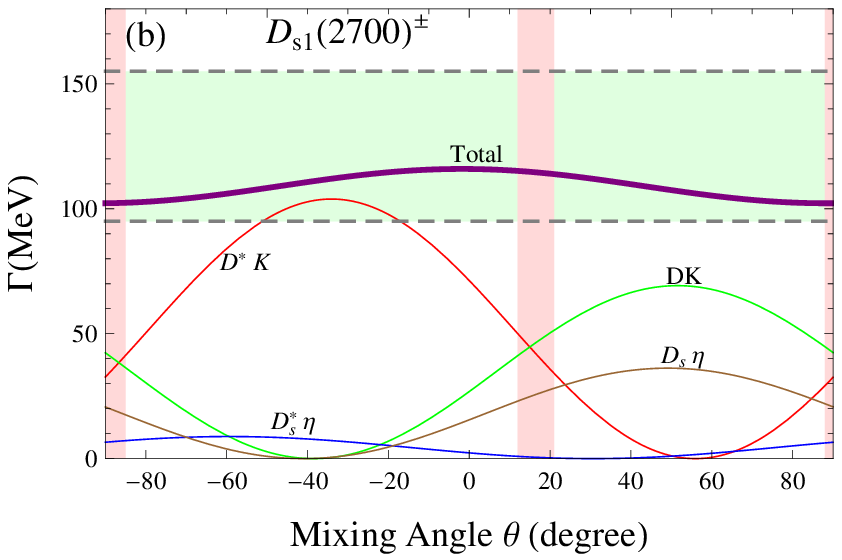}
\includegraphics[height=4.3cm]{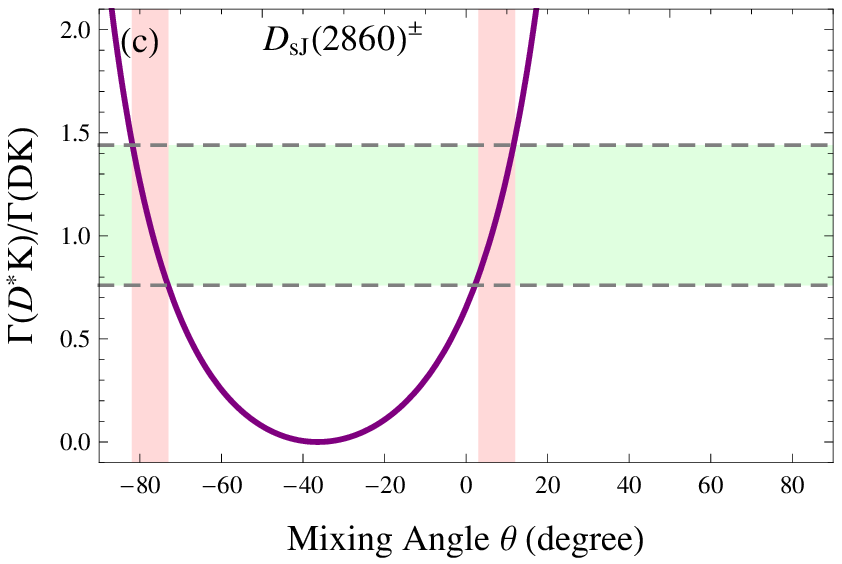}
\includegraphics[height=4.3cm]{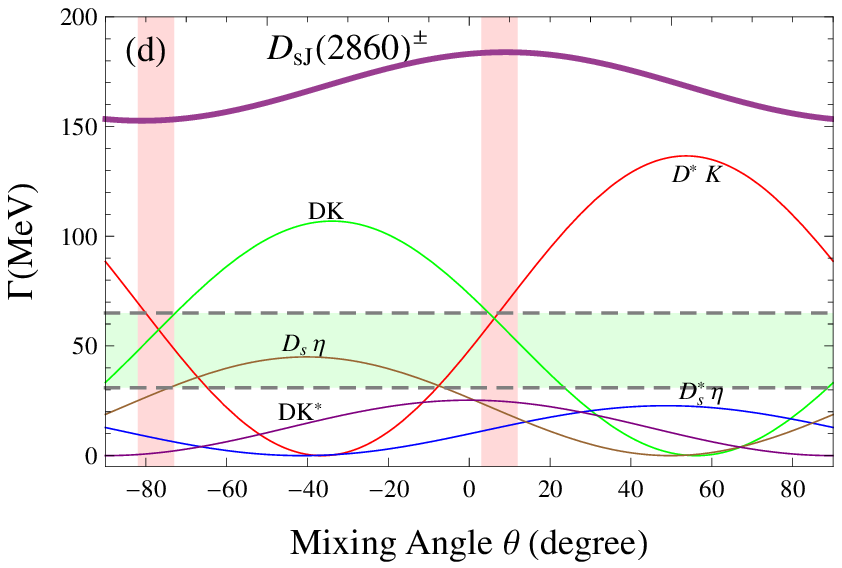}
\caption{Partial widths and $\Gamma(D^{\star}K)/\Gamma(DK)$ for the $D^\star_{s1}(2700)^\pm$ and $D^\star_{sJ}(2860)^\pm$ versus $\theta$ at $\beta=0.38$ GeV. The horizontal dashed lines indicate the upper and lower limits of the PDG data.}
\end{center}
\end{figure}

\subsection{Uncertainties within the $^3P_0$ model}

The uncertainties within the $^3P_0$ model have been studied for a long time~\cite{kokoski,geiger}. One uncertainty results from the form of the meson wave function. There are usually two choices for the wave functions. One choice is the SHO wave function and the other choice is the exact wave function resulted from the Coulomb-plus-linear potential. However, it is found out that the results are not strongly tied to a particular choice of wave functions~\cite{geiger,kokoski}. In our evaluation, the SHO wave functions are employed. In order to study the uncertainties, some relevant SHO wave functions are given explicitly in the momentum-space as
\begin{eqnarray}\label{br1}
{\Psi}_{\emph{nL$M_L$}}&=&\frac{1}{\beta^{\frac{3}{2}}} \left[ \frac{2^{l+2-n}(2l+2n+1)!!}{\sqrt{\pi} n! [(2l+1)!!]^2}
\right]^{\frac{1}{2}}  \nonumber \\
& &\times(\frac{k}{\beta})^l\exp \left[ - \frac{1}{2} (\frac{k}{\beta})^2 \right] \nonumber\\
& & \times F (-n,l+3/2,(\frac{k}{\beta})^2)Y_{\emph{L$M_L$}}(\Omega_\emph{p}),
\end{eqnarray}
where $\beta$ is the harmonic oscillator strength parameter, $Y_\emph{{L$M_L$}}(\Omega_\emph{p})$ is the spherical harmonic function, and $F (-n,l+3/2,(\frac{k}{\beta})^2)$ is the confluent hypergeometric function. Some relevant meson wave functions in the evaluation are listed as follows:
\begin{equation}\label{br3}
{\Psi}_{100}=\frac{1}{\beta^{\frac{3}{2}}\pi^{\frac{3}{4}}}\exp(-\frac{\textbf{k}^2}{2\beta^2}),
\end{equation}
\begin{equation}\label{br4}
{\Psi}_{200}=\sqrt{\frac{2}{3}}\frac{1}{\beta^{\frac{3}{2}}\pi^{\frac{3}{4}}}(\frac{\textbf{k}^{2}}
{\beta^{2}}-\frac{3}{2})\exp(-\frac{\textbf{k}^2}{2\beta^2}),
\end{equation}
\begin{equation}\label{br5}
{\Psi}_{12M_L}=\frac{4}{\sqrt{15}}\frac{1}{\beta^{\frac{7}{2}}\pi^{\frac{1}{4}}}Y_{12M_L}
\exp(-\frac{\textbf{k}^2}{2\beta^2}),
\end{equation}
where \textbf{k} is the relative momentum between a quark and an antiquark in a meson. For meson A composed of a quark $q_1$ and an antiquark $\overline{q}_2$,
\begin{equation}\label{br2}
\textbf{k}_A=\frac{m_2\textbf{k}_1-m_1\textbf{k}_2}{m_1+m_2},
\end{equation}
where $m_1$ and $m_2$ are the constituent quarks masses of quark $q_1$ and antiquark $\overline{q}_2$, respectively.

The constituent quark masses may also bring uncertainty to the strong decay widths. In different constituent quark models, the constituent quark mass may be different. For the strong decays of $D_s$, the heavy c quark and light $\overline {s}$ quark are involved in the initial state. For convenience, two dimensionless variables $\mu_1=m/m_1$ and $\mu_2=m/m_2$ are introduced, where $m$ is the mass of the light constituent quark q and antiquark $\overline{q}$ (q=u, d, s) created from vacuum, and $m_1$ and $m_2$ be the constituent quarks mass of c($\overline {c}$) and $\overline {s}$(s), respectively. To find out the explicit dependence of the total decay width of $D^\star_{s1}(2700)^\pm$ and $D^\star_{sJ}(2860)^\pm$ on these two $\mu_1$ ($=m_u/m_c$ or $m_s/m_c$) and $\mu_2$ (=$m_u/m_s$ or 1), these two variables could be thought as free variables in suitable region. In Fig. 3, the dependence of the total decay width on $\mu_1$ and $\mu_2$ (in a much larger region from $0$ to $1$) is shown. It is found that the total decay width depends weakly on $\mu_1$ and $\mu_2$. Similarly, the variation of the constituent quark masses affect the partial decay widths small.
\begin{figure}
\begin{center}
\includegraphics[height=4cm]{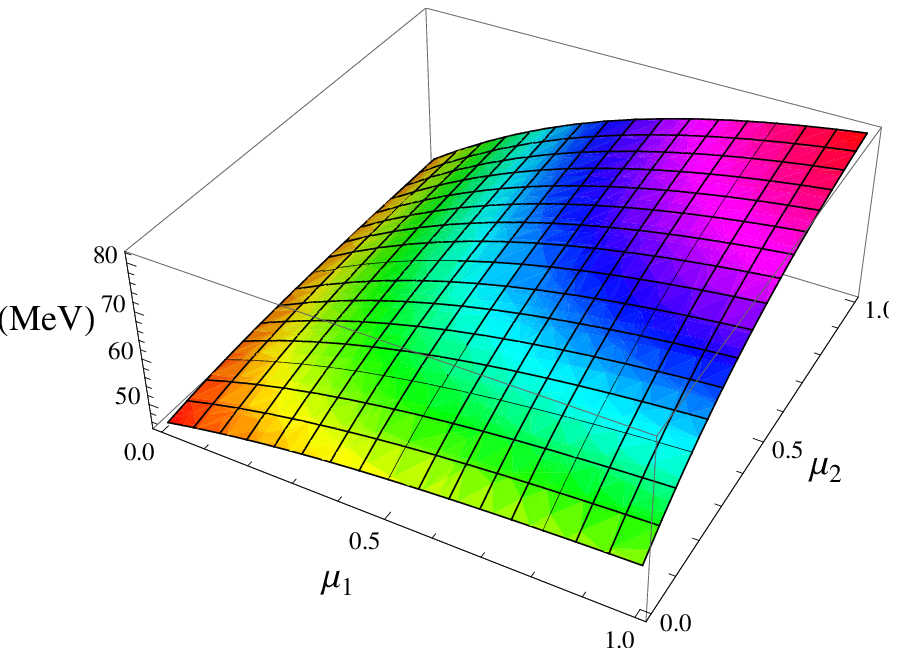}
\includegraphics[height=4cm]{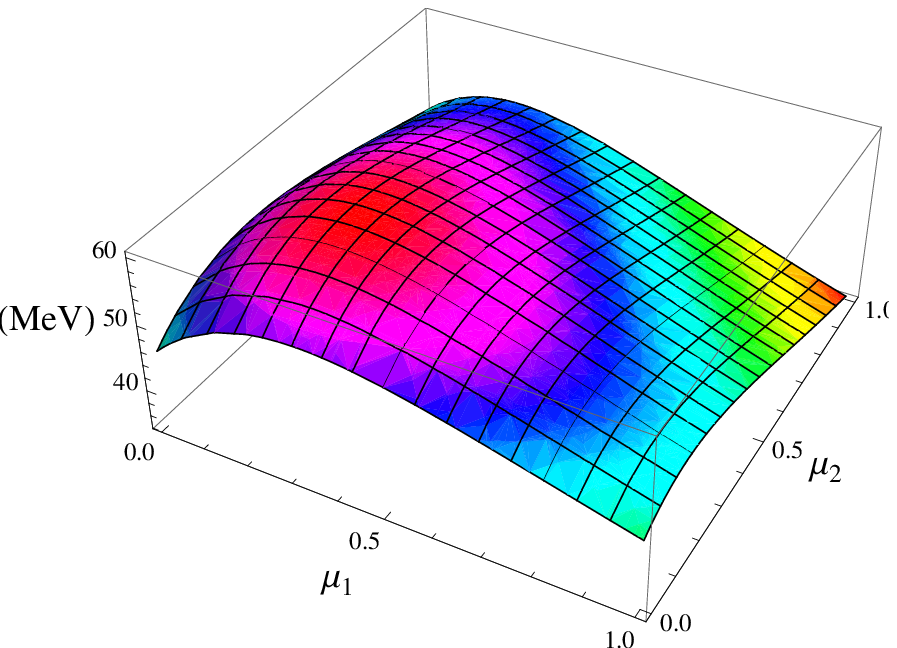}
\caption{Total decay widths of $D^\star_{s1}(2700)^\pm$ and $D^\star_{sJ}(2860)^\pm$ versus $\mu_1$ and $\mu_2$}
\end{center}
\end{figure}

Another uncertainty is related to the choices of phase space (PS)~\cite{geiger,isgur}. There are usually three choices for the phase space. The first way is to use the non-relativistic phase space, $PS=\emph{M}_\emph{B}\emph{M}_\emph{C}/\emph{M}_\emph{A}$. The second choice is the relativistic phase space, $PS=\emph{E}_\emph{B}\emph{E}_\emph{C}/\emph{M}_\emph{A}$. The third choice is the ``mock" phase space, $PS=\widetilde{\emph{M}}_\emph{B}\widetilde{\emph{M}}_\emph{C}/\widetilde{\emph{M}}_\emph{A}$, where $\emph{M}_\emph{i}$ is the mock mass of the state. The relativistic phase space is employed in previous evaluation. In Table. VI and Table. V, we present our results based on the non-relativistic phase space and the ``mock" phase space, respectively. From Table. II, III and IV, different choices of the phase space bring large uncertainty to the decay widths and branching ratio of $D^\star_{s1}(2700)^\pm$ as the $2^3S_1$. Different choices of the phase space bring large uncertainty to the decay widths but small uncertainty to the branching ratio of $D^\star_{s1}(2700)^\pm$ as the $1^3D_1$. Different choices of the phase space bring small uncertainty to the decay widths and branching ratio of $D^\star_{sJ}(2860)^\pm$ as the $1^3D_3$.

\begin{table*}
\begin{center}
\begin{tabular}{ccccccccc}
\hline\hline
\backslashbox{Assignment}{Mode}
& $D^{\star}K$ & $DK$ & $D^\star_s\eta$ & $D_s\eta$ &$DK^{\star}$ & $D_s\omega$ &$\Gamma_{totall}$ & $\Gamma(D^{\star}K)/\Gamma(DK)$\\
\hline
$D_{s1}(2700)[2^3S_1]$   & 30.4   & 5.8    & 1.9    & 1.5    & -    & -    & 39.6   & 5.3\\
$D_{s1}(2700)[1^3D_1]$   & 28.7   & 57.8   &1.9     & 12.8   & -    & -    & 101.2    & 0.5\\
$D_{sJ}(2860)[1^3D_3]$   & 14.0   & 17.5   & 0.5    & 1.2    & 1.9  &0.1   & 35.2   & 0.8\\
\hline\hline
\end{tabular}
\caption{Decay widths (MeV) and branching fraction ratios for $D^\star_{s1}(2700)^\pm$ and $D^\star_{sJ}(2860)^\pm$} \label{table-2}
\end{center}
\end{table*}

\begin{table*}
\begin{center}
\begin{tabular}{ccccccccc}
\hline\hline
\backslashbox{Assignment}{Mode}
& $D^{\star}K$ & $DK$ & $D^\star_s\eta$ & $D_s\eta$ &$DK^{\star}$ & $D_s\omega$ &$\Gamma_{totall}$ & $\Gamma(D^{\star}K)/\Gamma(DK)$\\
\hline
$D_{s1}(2700)[2^3S_1]$   & 48.5   & 10.2   & 3.1   & 2.7    & -    & -    & 64.4   & 4.8\\
$D_{s1}(2700)[1^3D_1]$   & 44.5   & 93.8   &2.8    & 21.9   & -    & -    & 163    & 0.47\\
$D_{sJ}(2860)[1^3D_3]$   & 21.9   & 30.5   & 0.8   & 2.3    & 1.9  &0.1   & 57.5   & 0.72\\
\hline\hline
\end{tabular}
\caption{Decay widths (MeV) and branching fraction ratios for $D^\star_{s1}(2700)^\pm$ and $D^\star_{sJ}(2860)^\pm$} \label{table-2}
\end{center}
\end{table*}

Finally, the uncertainty related the variation of $\beta$ as a universal parameter for all mesons is explored. To see the dependence of the results on $\beta$, $D^\star_{s1}(2700)^\pm$ and $D^\star_{sJ}(2860)^\pm$ are assumed as the pure states. The branching ratios $\Gamma(D^{\star}K)/\Gamma(DK)$ and the total decay widths of $D^\star_{s1}(2700)^\pm$ and $D^\star_{sJ}(2860)^\pm$ are plotted in Fig. 4. From this figure, the variation of $\beta$ in the preferred region brings very small uncertainty to the total decay width of $D^\star_{s1}(2700)^\pm$, but may bring $30\%$ uncertainty to the total decay width of $D^\star_{sJ}(2860)^\pm$. The variation of $\beta$ bring considerable uncertainties to the branching ratios of $D^\star_{s1}(2700)^\pm$, but smaller uncertainty to the branching ratios of $D^\star_{sJ}(2860)^\pm$.

\begin{figure} \label{pic3}
\begin{center}
\includegraphics[height=4.7cm]{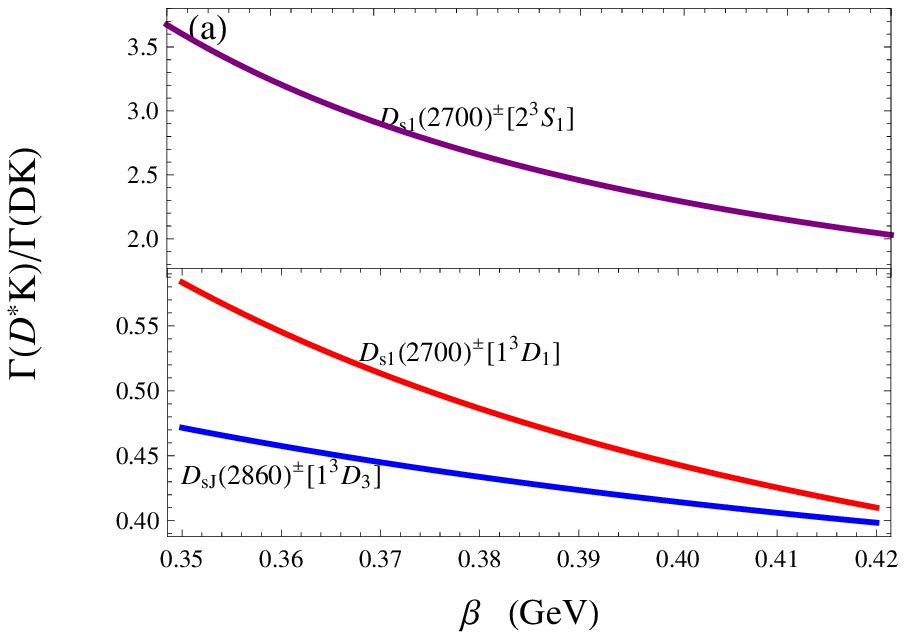}
\includegraphics[height=4.3cm]{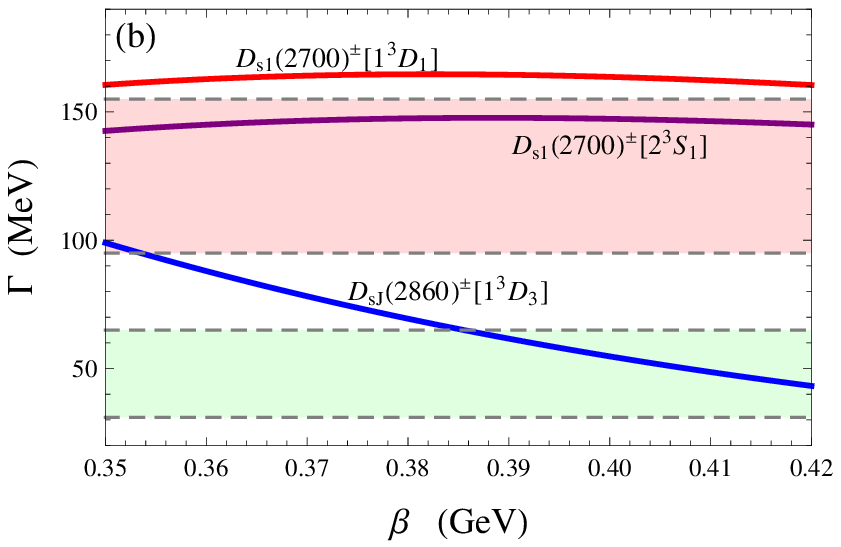}
\caption{Branching ratios and total decay widths of $D^\star_{s1}(2700)^\pm$ as $2^2S_1$ and $D^\star_{sJ}(2860)^\pm$ as $1^3D_3$ versus $\beta$}
\end{center}
\end{figure}

\section{Conclusions and discussions}

In this work, we have studied the strong decay properties of $D^\star_{s1}(2700)^\pm$ and $D^\star_{sJ}(2860)^\pm$ in the $^3P_0$ model. It is found that the interpretation of these two states depends on the mixing schemes and the ways of choices of the harmonic oscillator parameter $\beta$. In the evaluation of the strong decays widths, there are often two ways to choose the the harmonic oscillator parameter $\beta$. One way is to choose $\beta$ individually for each meson, and the other way is to choose $\beta$ universal for all mesons. The radially excited $2^3S_1$ and the orbitally excited $1^3D_1$ may mix with each other. The interpretations of $D^\star_{s1}(2700)^\pm$ and $D^\star_{sJ}(2860)^\pm$ depend on whether the mixing exists or not. When the mixing exists, the interpretations of $D^\star_{s1}(2700)^\pm$ and $D^\star_{sJ}(2860)^\pm$ depend on these two different ways of choices of $\beta$.

In the case of special $\beta$ for each meson, $D^\star_{s1}(2700)^\pm$ can not be interpreted as the pure $2^3S_1$ $D_s$, but may be interpreted as the $1^3D_1$. $D^\star_{sJ}(2860)^\pm$ may be interpreted as the pure $1^3D_3$. When the mixing scheme is employed, $D^\star_{s1}(2700)^\pm$ prefers to be identified with the mixed $1^-$ state of $2^3S_1$ and $1~^3D_1$, and $D^\star_{sJ}(2860)^\pm$ prefers to be identified with the orthogonal partner of $D^\star_{s1}(2700)^\pm$ with a large mixing angle $\theta$, which implies that $1^3D_1$ is predominant.

In the case of fixed $\beta$ for all mesons, $D^\star_{s1}(2700)^\pm$ can not be interpreted as a pure $2^3S_1$, but may be interpreted as the $1^3D_1$. $D^\star_{sJ}(2860)^\pm$ may be interpreted as the pure $1^3D_3$. $D^\star_{s1}(2700)^\pm$ can also be interpreted as the mixed state of $2^3S_1$ and $1^3D_3$ with a small mixing angle $\theta$ ($2^3S_1$ is predominant). However, it is hard to interpret $D^\star_{sJ}(2860)^\pm$ as the orthogonal partner of $D^\star_{s1}(2700)^\pm$.

Without mixing between the orbitally and radially excited states, the interpretations of $D^\star_{s1}(2700)^\pm$ and $D^\star_{sJ}(2860)^\pm$ are the same for all $\beta$ no matter $\beta$ is universal or not. However, the conclusions are different for different ways of choices of $\beta$ when the mixing exists. This uncertainty my be inherent within the $^3P_0$ model and appears accidently when the mixing exists. This uncertainty blurs our understanding of the observed states. To draw conclusions for $D^\star_{s1}(2700)^\pm$ and $D^\star_{sJ}(2860)^\pm$ through their strong decays within the $^3P_0$ model, one must be careful when the mixing exists. The analysis of the strong decays in other models is necessary. Of course, the study of the production and other ways of decays will provide other important understandings to them. Some other uncertainties within the $^3P_0$ model are explored. The constituent quark masses bring small uncertainty to the results, but the phase spaces may bring considerable one.
\section*{ACKNOWLEDGMENTS}
This work is supported by the National Natural Science Foundation of
China under Grant No. 11075102. Bing Chen is also supported by Shanghai University under the Graduates Innovation
Fund SHUCX111023 (No. A. 16-0101-10-018).

\end{document}